\RequirePackage{ifpdf}
\documentclass{PoS}

\graphicspath{{figs/}} 

\usepackage{amsfonts, amssymb, amsmath}

\usepackage{color}
\usepackage{caption, subcaption}
\let\normalcolor\relax

\newcommand{\Dmix}{\Delta_\text{mix}}

\title{Hadron spectra and $\Dmix$ from overlap quarks\\ on a HISQ sea}

\ShortTitle{Spectra \& $\Dmix$ from overlap on HISQ.}

\author{S.\ Basak \\
School of Physical Sciences, 
National Institute of Science Education and Research,
Bhubaneswar 751005, India
}

\author{S.\ Datta, N.\ Mathur \\ 
Dept.\ of Theoretical Physics,
Tata Institute of Fundamental Research,
Mumbai 40005, India 
}

\author{\speaker{A.T.\ Lytle} \\ 
Dept.\ of Theoretical Physics,
Tata Institute of Fundamental Research,
Mumbai 40005, India \\
SUPA, 
School of Physics and Astronomy, 
 University of Glasgow, 
 Glasgow G12 8QQ, UK 
 }

\author{P. Majumdar \\
Department of Theoretical Physics, Indian Association for the Cultivation of Science, Kolkata
700032, India.
}

\author{M. Padmanath\\
Institute of Physics, University of Graz, 8010 Graz, Austria.
}

\author{ILGTI Collaboration}
\abstract{
We present results of our continuing study on mixed-action hadron spectra and decay constants
using overlap valence quarks on MILC's 2+1+1 flavor HISQ gauge configurations.
This study is carried out on three lattice spacings,
with charm and strange masses tuned to their physical values, and with $m_l/m_s = 1/5$.
We present results of an ongoing determination of the mixed-action parameter $\Dmix$, which enters into chiral formulae for the masses and decay constants.
}

\FullConference{The 32nd International Symposium on Lattice Field Theory\\
                 23 - 28 June,  2014\\
                 Columbia University New York, NY}
                 
\begin{document}

\section{Introduction }  \label{sec:intro}
There has been a resurgence of interest in heavy hadron spectroscopy,
with recent discoveries of numerous hadrons with one or more heavy quarks.
Including heavy quarks in lattice QCD simulations remains challenging,
since for present-day simulations $am_h \ll 1$ is in general not satisfied.

We have adopted a mixed-action approach using 
the overlap fermion action~\cite{Neuberger:1997fp}
on a 2+1+1 flavor HISQ sea~\cite{Follana:2006rc}.
The overlap action is $\mathcal{O}(am)$ improved; one aim of the present study is
to investigate its behavior in the regime $am \lesssim 1$.
Because it maintains chiral symmetry, the analysis of many lattice observables is simplified when using the overlap action.  
However the generation of dynamical fermions with this action is prohibitively costly.  Instead
we use the 2+1+1 flavor highly-improved staggered (HISQ) configurations
made available by the MILC collaboration~\cite{Bazavov:2012xda}.

Here we present the current status of our calculation of hadrons with charm,
and we also calculate the combination $a^2 (\Dmix + \Dmix')$ on the coarse MILC ensemble ($a \sim 0.12$ fm),
which determines the lattice-spacing dependent shift 
in the mass of valence-sea pions due to using a mixed action~\cite{Chen:2009su}, similar to the quantities
$\Delta_t$ that parametrize taste-breaking.
We estimate the mixed-action parameter $\Dmix$, related to the mixed-action low energy constant $C_{\text{mix}}$ by $\Dmix = 16 \, C_{\text{mix}} / f^2$.
A similar mixed-action approach with overlap valence on 2+1 flavor
dynamical domain-wall configurations has been used by the $\chi$QCD
collaboration~\cite{Li:2010pw,Mathur:2010ed},

\section{Simulation details}
We present results from three ensembles of 2+1+1 flavor dynamical HISQ
fermions, generated by the MILC collaboration.
These ensembles have extents of $48^3 \times 144$, $32^3 \times 96$,
and  $24^3 \times 64$ with $10/g^2 = 6.72$, 6.30 and 6.00, respectively.
In all cases the charm and strange masses are tuned to near their physical
values, while $m_l/m_s$ is fixed to $1/5$.
Results on the $32^3$ and $48^3$ ensembles 
were presented in~\cite{Basak:2013oya,Basak:2012py}.

We have independently determined the lattice spacings
by equating the lattice determined $\Omega_{sss}$ mass with its physical
value.  
Here the valence strange mass is determined by setting the $\bar{s} s$
mass to 685 MeV~\cite{Davies:2009tsa}. 
We obtained
lattice spacings of $0.0582(5)$, $0.0877(10)$ and $0.1192(14)$ fm, respectively, for finer to coarser lattices, 
which are consistent with the values $0.0582(5)$, $0.0888(8)$,  and $0.1207(11)$ fm 
obtained by the MILC collaboration using the $r_1$ parameter~\cite{Bazavov:2012xda}.

Valence quark propagators are computed using the overlap action.
The numerical implementation follows the methods used by the $\chi$QCD
collaboration~\cite{Mathur:2004jr,Chen:2003im} as discussed in ~\cite{Basak:2013oya,Basak:2012py}. 
We use periodic/antiperiodic boundary conditions in space/time.  
The gauge fields are fixed to Coulomb gauge and smeared with a 
single HYP~\cite{Hasenfratz:2001hp} blocking transformation.
We use point-point, wall-point, and wall-wall propagators to calculate the hadron correlation functions.

The valence charm mass is tuned 
by setting the spin-averaged 1S state mass 
$(m_{\eta_c} + 3 m_{J/ \psi})/4$ to its physical value,
using the kinetic mass obtained from the 1S dispersion 
relation. Using the kinetic mass we
find a value of $c$ much closer to 1 as compared to the rest
mass. This is discussed further in ~\cite{Basak:2013oya,Basak:2012py}.

\section{Charmed hadron spectrum}  \label{sec:spect}
\begin{figure}
\centering
\includegraphics[width=0.65\textwidth]{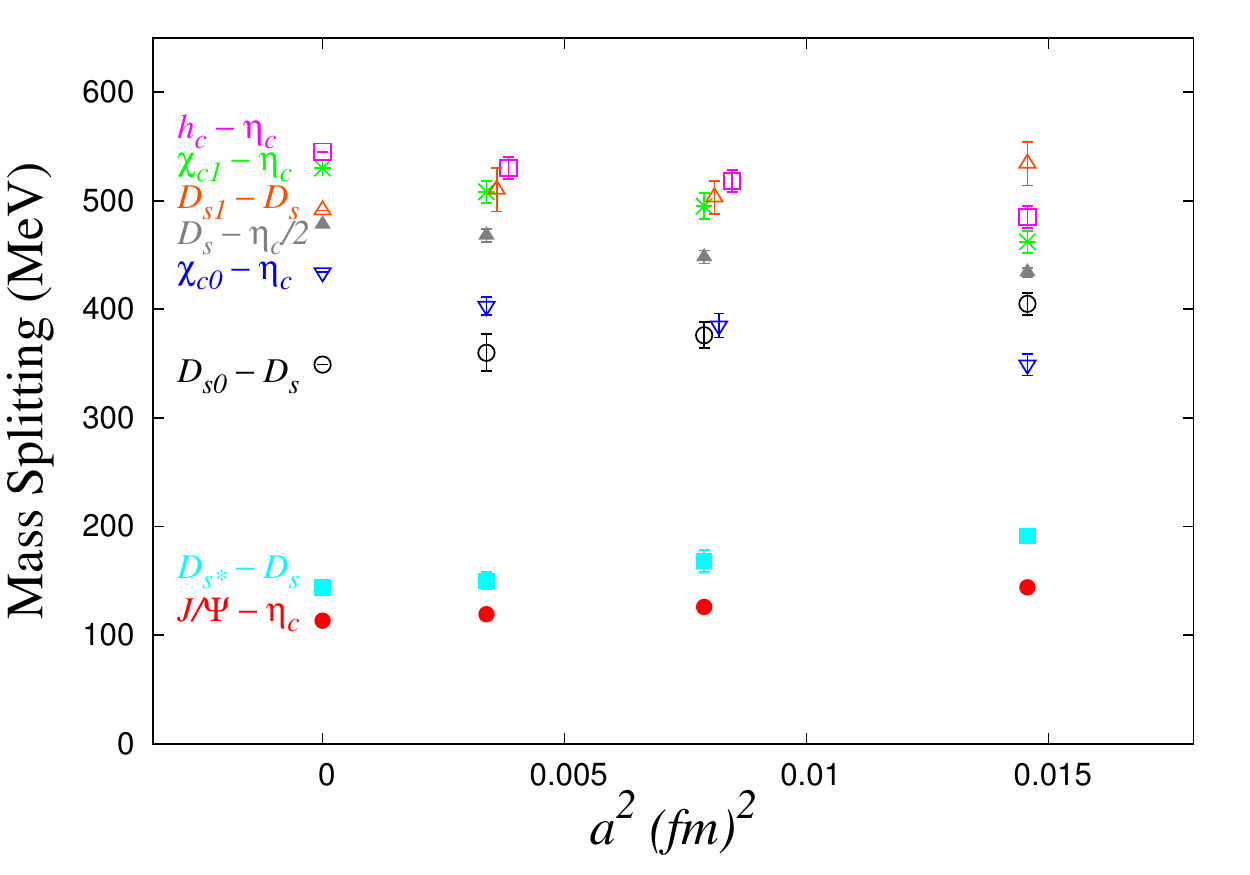}
\caption{Meson mass splitting for charmonia and charmed-strange mesons at
three lattice spacings. Experimental values are shown in the left side. 
\label{fig1}}
\end{figure}

\begin{figure}
\centering
\includegraphics[width=0.65\textwidth]{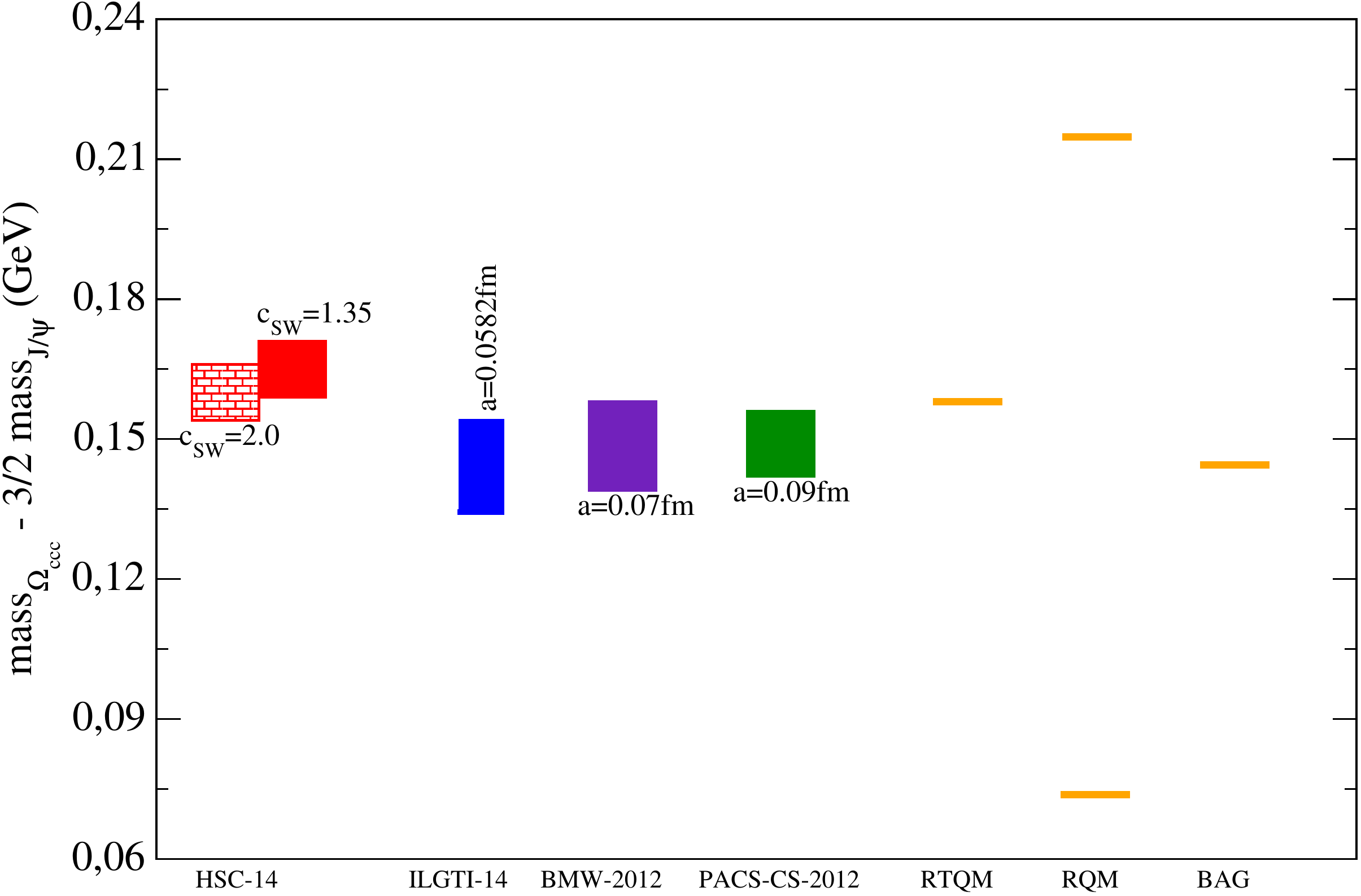}
\caption{The mass splitting $m_{\Omega_{ccc}} - {3\over2} m_{J/\Psi}$ along with other 
lattice and various model results. Result from this work is shown in blue color.
 References in the plot are (from left to right)
  HSC-14~\cite{Padmanath:2013zfa},  ILGTI-14 this work, BMW-2012~\cite{Durr:2012dw},
 PACS-CS-2012~\cite{Namekawa:2013vu}, RTQM~\cite{Martynenko:2007je}, 
 RQM~\cite{Migura:2006ep}, BAG~\cite{Hasenfratz:1980ka}.
\label{fig1b}}
\end{figure}

Figure~\ref{fig1} shows our results for charmonia and charmed-strange
mesons at three lattice spacings. Results
are presented in terms of splittings with respect to $\eta_c$ and
$D_s$ mesons, obtained by fitting directly the ratios of the correlators. We have
not performed any continuum or chiral extrapolation yet and the numbers at the continuum limit are taken from PDG. However, it
is to be noted that there is a clear tendency of convergence of our
lattice results with physical values in the continuum limit.  It is
expected that the discretization error will be maximum for
triply-charmed baryons because of the presence of three heavy charm
quarks. In Fig.~\ref{fig1b} we plot our results for the ground state energy of the spin-3/2
triply-charmed baryon minus 3/2 times the $J/\Psi$ mass. The 3/2
factor is included to cancel out the effect of charm quarks.  Our results are
shown in blue along with other lattice results. A
few model results are also shown on the right. In Fig.~\ref{fig2}(a) we show the
hyperfine splitting of 1S charmonia at three lattice spacings along
with its physical value. It is to be noted that the continuum limit
value of this splitting for overlap fermions, utilized in this work,
is approached from above. This is in contrast to the result obtained for this 
quantity using Wilson valence fermions where it is approached from below~\cite{DeTar:2012xk}
 and for HISQ fermions where it is approached from above for coarser lattices and from
below for finer lattices~\cite{Donald:2012ga}. 
We have not performed continuum and chiral
extrapolation here though a naive fit with a form $\delta_{phys} = A +
B a^2$ gives a value 110(4) MeV which is consistent with the physical value. 
In Fig.~\ref{fig2}(b) we show
mass splittings $m_{\Omega_{ccc}} - {3\over2} m_{J/\Psi}$ at three lattice
spacings.  Again a naive fit with a form $\delta_{phys} = A + B a^2$
yields a value of this splitting as 148(10) MeV.  This splitting
should be comparable to 
the binding energy of the yet to be discovered spin-3/2 triply
charmed baryon.

\newcommand{\figspec}[3]{
\includegraphics[width=#1\textwidth,height=#2\textwidth,clip=true]{#3}
}

\begin{figure}[h] 
\figspec{0.5}{0.32}{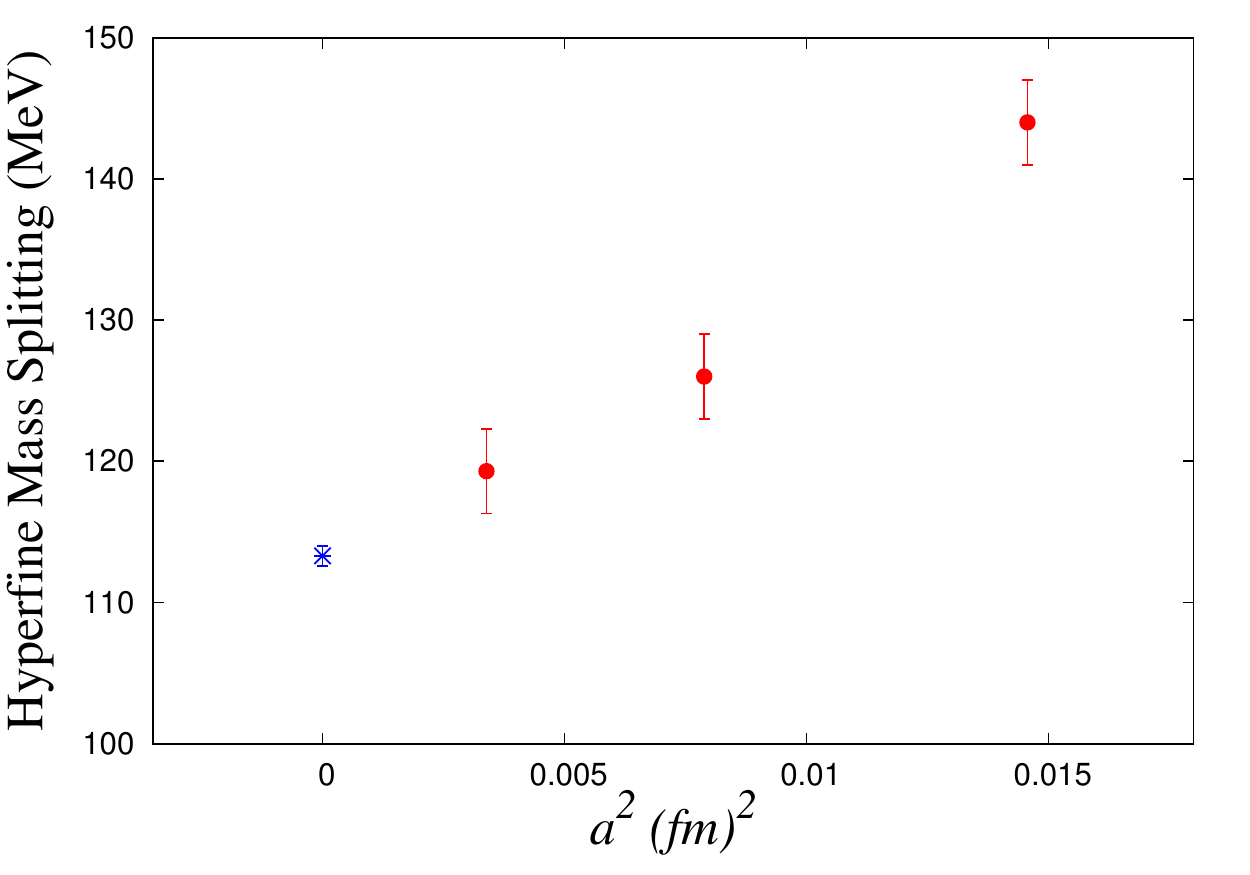}
\figspec{0.5}{0.32}{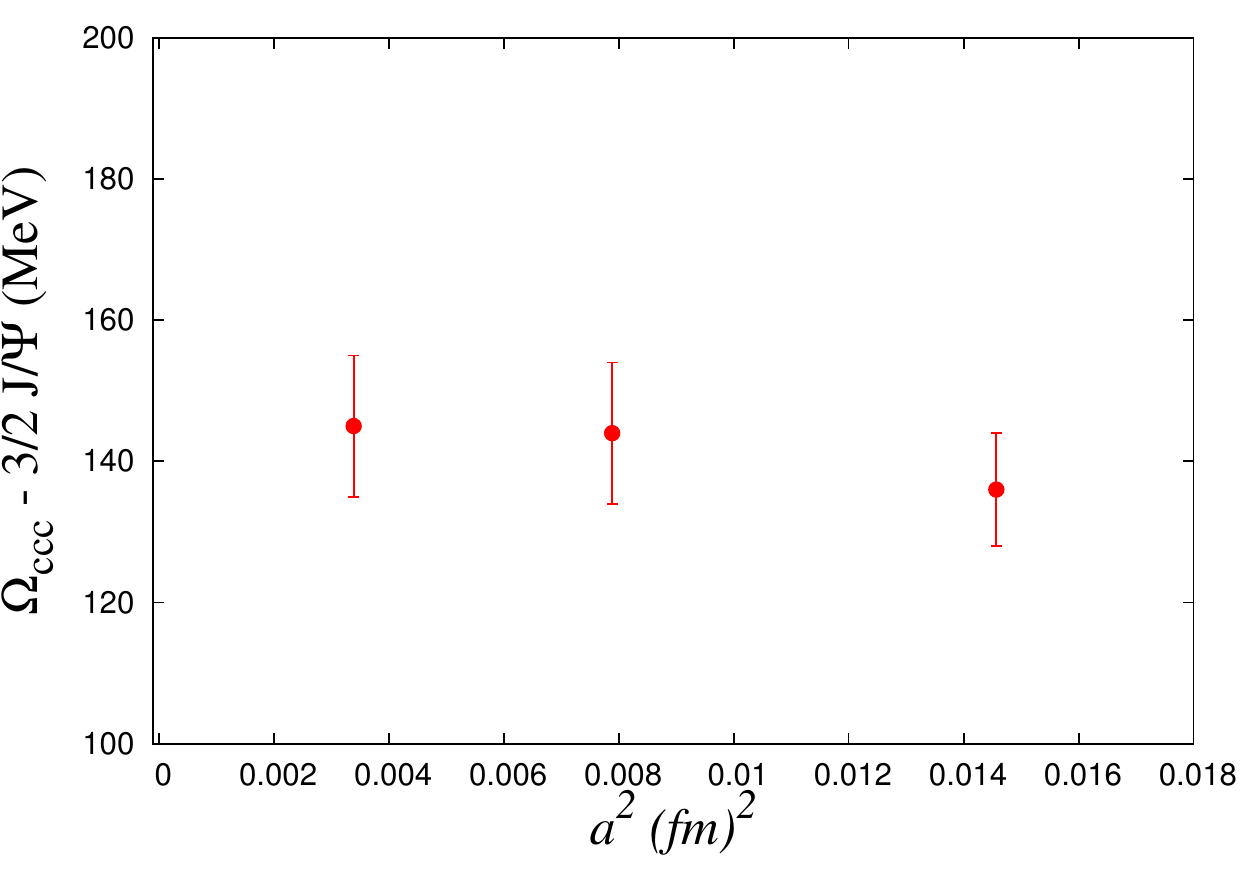}
\caption{(a) Hyperfine mass splitting of 1S charmonia at three lattice spacings (red circles) along with its physical value (blue star). (b) The mass splitting $m_{\Omega_{ccc}} - {3\over2} m_{J/\Psi}$ at three lattice spacings.
\label{fig2}}
\end{figure} 

\section{$\Dmix$ for overlap on HISQ }  \label{sec:dmix}
The low energy properties of a simulation
employing different sea and valence actions
can be described near the chiral limit using 
mixed-action chiral perturbation theory (MA$\chi$PT)~\cite{Bar:2002nr}.
This formalism extends the usual $\chi$PT description
by terms that are proportional to new low-energy constants,
and vanish in the continuum limit.
For staggered simulations, $\chi$PT is extended
by terms describing taste-breaking discretization effects,
yielding staggered chiral perturbation theory (S$\chi$PT)~\cite{Lee:1999zxa}.
For chiral valence fermions on a staggered sea, it has been shown
that only one new low-energy constant, $C_{\text{mix}}$, appears at leading order
in addition to those arising in S$\chi$PT~\cite{Bar:2005tu}.
Here we estimate the parameter $\Dmix = 16 \, C_{\text{mix}}/f^2$ 
for overlap fermions on the coarse HISQ ensembles.

Several studies in recent years have studied the size of these
effects in the context of staggered sea fermions.
Domain-wall valence fermions on the 
MILC collaboration's asqtad~\cite{Bazavov:2009bb} ensembles
were studied in~\cite{Orginos:2007tw,Aubin:2008wk}.
In Refs.~\cite{Chen:2009su,Aubin:2008wk}
it was pointed out that the quantity $a^2 (\Dmix + \Dmix')$ is comparable
in magnitude to the size of mass splittings between pions of different tastes.
One of the primary advantages of the HISQ action 
is the reduced taste-symmetry violations~\cite{Bazavov:2012xda}.  
We expect a comparable
reduction in this quantity when using chiral fermions on the
HISQ ensembles,
and find this to be the case.

At the leading order in MA(S)$\chi$PT, 
the masses of pions constructed from 
valence~($v$) and sea~($s$)
action propagators are given by
\begin{align}
m^2_{vv'} &= B_{\text{ov}} (m_v + m_{v'})  \label{vv'} \\
m^2_{ss'} &= B_{\text{HISQ}}(m_s + m_{s'}) + a^2 \Delta_{t} \label{ss'}\\
m^2_{vs} &= B_{\text{ov}} m_v + B_{\text{HISQ}} m_s + a^2 (\Dmix + \Dmix') \,.  \label{vs}
\end{align}
Eq.~\eqref{ss'} gives the well-known splitting of different taste pions
in terms of $\Delta_t$, while pions constructed from one valence-action
propagator and one sea-action propagator on the sea-action ensemble
are sensitive to $\Dmix + \Dmix'$.  $\Dmix$ is related to the MA$\chi$PT LEC
by $\Dmix = 16 \, C_{\text{mix}}/f^2$, 
while $\Dmix'$ is given in terms of staggered taste splittings~\cite{Chen:2009su}.


Different strategies for extracting $\Dmix$ from Eqs.~\eqref{vv'}-\eqref{vs} have
been proposed in the literature.  Here we adopt the same technique as
in~\cite{Aubin:2008wk,Lujan:2012wg}, fitting the quantity
\begin{equation}
\delta m^2(m_v) \equiv m^2_{vs} - m^2_{ss}/2 = B_{\text{ov}} m_v + a^2 (\Dmix + \Dmix')
\end{equation}
as a function of valence quark mass.  This is convenient since it is 
the valence propagators that are used in e.g. spectroscopy calculations.

The mixed-meson correlation functions are constructed using one
overlap propagator and one \emph{Wilsonized} staggered propagator.
The Wilsonized propagator $G_{\psi_s}$ is given in terms of the staggered
propagator $G_\chi$ by~\cite{Wingate:2002fh}
\begin{equation}
G_{\psi_s}(x,y) =  \Omega(x) \Omega^{\dagger}(y) \times G_{\chi}(x,y) \,,
\end{equation}
where $\Omega(x)$ is the Kawamoto-Smit transformation
\begin{equation}
\Omega(x) = \prod_\mu (\gamma_{\mu})^{x_\mu} \,.
\end{equation}

Mixed-meson correlators are fit to the form
\begin{equation}
C^{\Gamma}_{vs}(t) \sim [A + (-1)^{t} B] \cosh(m_{vs} (t-T/2)) \,.
\end{equation}

\begin{figure}
\begin{center}
\includegraphics[width=.65\textwidth]{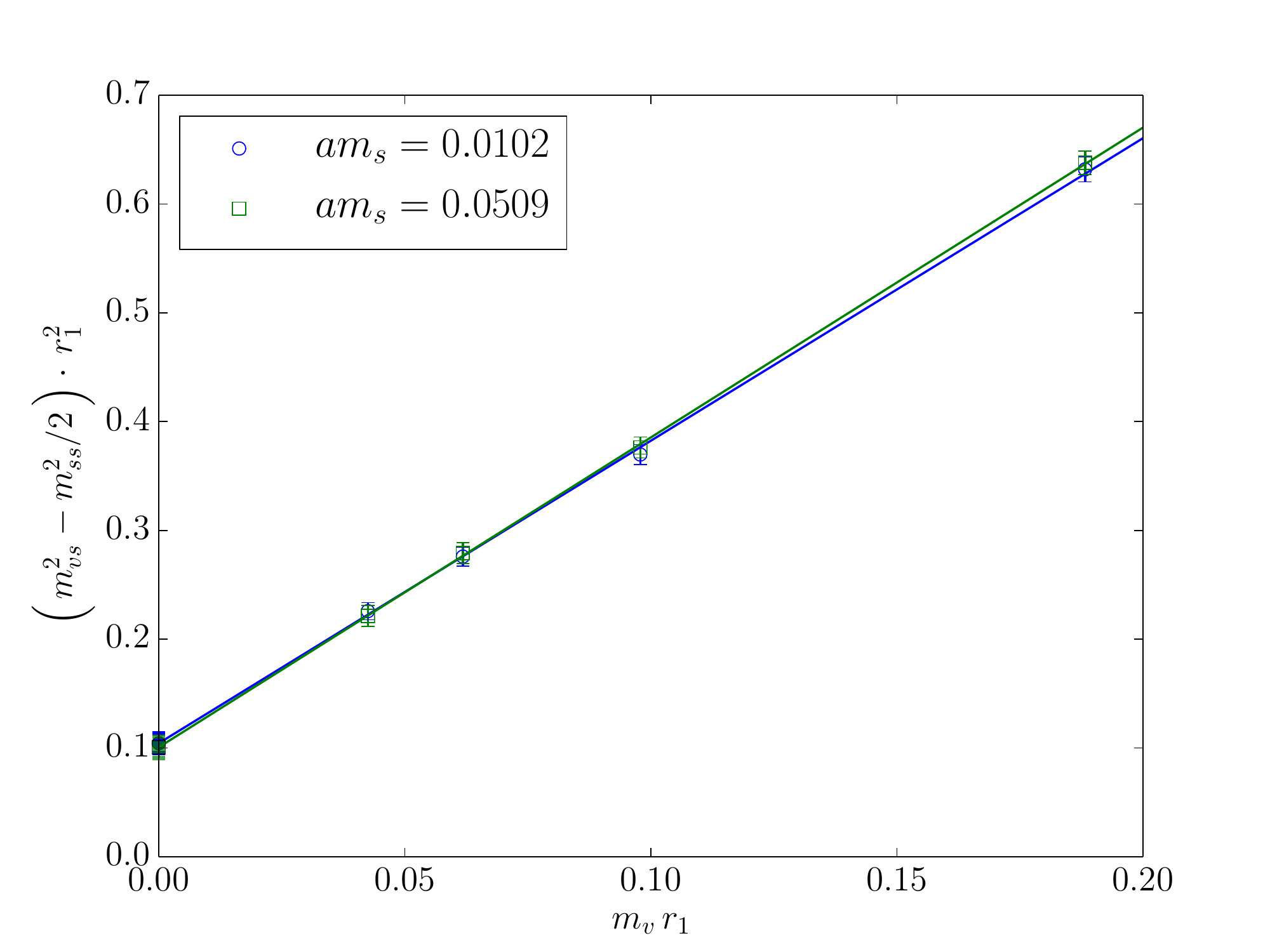}
\end{center}
\caption{$\delta m^2$ vs.\ $m_v$ in units of $r_1^{-1}$,
for two different values of the HISQ sea mass.
\label{fig: dmsq}}
\end{figure}

Figure~\ref{fig: dmsq} shows  
$\delta m^2$ vs.\ $m_v$ (in units of $r_1^{-1}$),
for a variety of valence quark masses and at two different values of sea mass $m_{s}$, 
on the coarse HISQ ensemble $a=0.121 \text{ fm}$.
The data is consistent with a straight line and
insensitive to the HISQ sea mass $m_s$, indicating that
Eqs.\ \eqref{vv'}--\eqref{vs} are valid for the ranges of quark masses used.
From the intercept of this data we find
\begin{equation}
r_1^2 a^2 (\Dmix + \Dmix') = 0.104(9) \,,
\end{equation}
and combining this with $a^2 \Dmix'$ as determined from the known taste splittings~\cite{Bazavov:2012xda},
we find
\begin{equation}  \label{dmix_val}
a^2 \Dmix \simeq (140 \text{ MeV})^2  \quad (a=0.121 \text{ fm}).
\end{equation}
In order to convert this into a continuum determination of $\Dmix$,
the calculation needs to be repeated at finer lattice spacings
and the results extrapolated to $a=0$. 

In~\cite{Aubin:2008wk} it was pointed out that the mixed-meson mass shift for domain-wall on asqtad
is comparable in size to the pion taste splittings.
Taste splittings for the HISQ action are reduced by a factor
$\gtrsim$ 3 relative to asqtad.
We find that for overlap on HISQ, the shift is comparable to the HISQ taste splitting,
and smaller than the asqtad taste splittings by around a factor of two.

\section{Conclusions }  \label{sec:concl}
We have presented an update of our results concerning charmed meson and baryon
spectroscopy using overlap valence fermions on the HISQ ensembles
made available by the MILC collaboration.
Such a mixed-action approach is attractive in that one gains the
advantages of the overlap Dirac operator in the valence sector while avoiding
the extreme cost of ensemble generation using such an operator.
One sensitive issue is whether simulating charm directly in such a setup
discretization errors can be under control.
Our studies of charm-meson dispersion relations employing the kinetic mass
indicate that this setup is suitable for charm spectroscopy at the current
lattice spacings.

The combination $a^2(\Dmix + \Dmix')$ provides a measure of mixed-action
effects in the chiral regime. 
We determined this combination for overlap fermions on a HISQ sea at a single lattice spacing,
finding it comparable in magnitude to
the pion taste-splittings,
and about half as large as was found for simulations of domain-wall on asqtad.
This calculation needs to be repeated at finer lattice spacings
to find a continuum result for $\Dmix$.

\section{Acknowledgements}
We thank M.\ Golterman for bringing Ref.~\cite{Chen:2009su} to our attention.
Computations were carried out on the Blue Gene/P 
of the Indian Lattice Gauge Theory Initiative,
Tata Institute of Fundamental Research (TIFR), Mumbai.
Part of this work is carried out in the cluster funded by DST-SERB project No. SR/S2/HEP-0025/2010.
We thank A.\ Salve and K.\ Ghadiali for technical support.
We are grateful to the MILC collaboration and in particular
to S. Gottlieb for providing us with the HISQ lattices.


\let\oldbibliography\thebibliography
\renewcommand{\thebibliography}[1]{%
  \oldbibliography{#1}%
  \setlength{\itemsep}{-1pt}%
}

\end{document}